\documentclass[12pt]{llncs}

\pdfoutput=1

\usepackage{epsfig}
\usepackage{amsmath}
\usepackage{amssymb}
\usepackage{subfigure}
\usepackage{fullpage}

\def\Proof{\par\noindent{\bf Proof:}\indent}
\def\QED{\hfill$\Box$\par\vskip1em}

\begin{document}

\title{Stateless Geocasting}
\author{Jordan Adamek\inst{1} \and Mikhail Nesterenko\inst{1} \and
        S\'{e}bastien Tixeuil\inst{2}\thanks{This work was supported in part by LINCS.}}
\institute{Kent State University \and UPMC Sorbonne Universit\'{e}s \& IUF
}

\maketitle

\thispagestyle{plain}
\pagestyle{plain}

\begin{abstract}
We present two stateless algorithms that guarantee to deliver the
message to every device in a designated geographic area: flooding and
planar geocasting. Due to the algorithms' statelessness, intermediate
devices do not have to keep message data between message
transmissions. We formally prove the algorithms correct, estimate
their message complexity and evaluate their performance through
simulation.
\end{abstract}

\section{Introduction}
\noindent
The advent of ubiquitous wireless networks, from sensor networks
tracking environmental patterns to metropolitan areas offering free
wireless Internet services to residents, upends classical means of
routing and delivering information. The scale, volatility and dynamic
nature of these networks present a formidable challenge.

One of the simplest routing algorithms for wireless networks is
\emph{controlled flooding} where each device retransmits the message
to all its neighbors. The device needs to store the information about
the transmission to prevent duplicate message resents. Flooding
potentially involves all communication devices of the network and,
therefore, does not scale well. Early routing algorithms are typically
routing-table based~\cite{JM96,PBD03}. However, maintenance of these
tables is resource intensive and is often infeasible.  Geometric
routing offers more scalable and resource frugal solutions to wireless
navigation. In \emph{geometric routing}, message forwarding decisions
are based on communication device coordinates. These may be physical
coordinates obtained, for example, from GPS, or virtual coordinates
computed by devices
themselves~\cite{IN99,KV99,KV00,KSU99,NI97}. Routing may be unicast,
where message is to be delivered to a single target device, or
multicast where there are several targets.

Geometric routing allows \emph{stateless} communication where devices
do not store any information about the transmitted message between
transmissions. This is a particularly attractive property: it scales
well since no multihop routing information need to be maintained by the
communicating devices; it is energy efficient since resources are not
spent on topology updates; and configuration change- and fault- tolerant
as the system trivially adjusts to them.  A number of unicast
stateless geometric routing algorithms is presented in the
literature~\cite{BMSU01,CMN12,DSW02,KK00,KV00,KSU99,KWZZ03}.

\emph{Geocasting} is a particular variant of multicasting where the
source device wishes to send the information to all devices located
in a specific geographic region. For example, geocasting may be used
to notify all households in  a flood-risk area once the water level
reaches some critical point; or to locate a moving vehicle whose last
known coordinates are available at the source. Geometric routing may
be particularly suitable for geocasting.

\ \\ \textbf{Related work.} Let us cover unicast geometric routing
first. The simplest form of geometric routing is greedy. In
\emph{greedy routing} each device selects the next hop neighbor with
the closest Euclidean distance to the target. However, greedy routing
fails if some device is the closest to the destination in its
immediate neighborhood. That is, this device is a \emph{local
  minimum}. Face routing guarantees delivery by navigating around
faces of a planarized communication graph \cite{BMSU01}. Face routing
may be inefficient if traversed faces are large.
Greedy-Face-Greedy~\cite{DSW02} starts in greedy mode and switches to
face routing only in case greedy fails. Once recovered, it switches
back to greedy. Face traversal may be inefficient if its traversal
direction is selected inopportunely: face traversal distance may be
long in one direction and short in the other.  GOAFR+\cite{KWZZ03} finds
the shorter traversal direction by switching it once the message
reaches a pre-determined ellipse containing source and target devices.
Concurrent Face Routing~\cite{CMN12} optimizes the speed of message
delivery by sending two concurrent messages in the opposite traversal
directions.

Let us now discuss existing geometric geocasting
algorithms. Geographic-Forwarding-Geocast~\cite{SH06} starts as a
geometric unicast until it reaches the geocast region. Once inside the
region, the message is flooded. The flood messages that reach devices
outside the geocast region are discarded. The flooding is
stateful. Moreover, as noted by Casteigts et al.~\cite{CNS10},
Geographic-Forwarding-Geocast may fail to deliver the message to all
devices in the geocast region, if the region is disconnected and the
only connectivity is through outside devices. Virtual Surrounding Face
\cite{LNLC06} avoids this problem by pre-computing in advance all
planar faces that intersect the geocast region. The algorithm unicasts
to the region and, upon reaching the geocast area, floods it inside
and traverses all the precomputed surrounding faces on the
outside. The algorithm is stateful. Also, the pre-computation and
maintenance of the virtual surrounding face is by nature stateful and
may incur significant overhead in a dynamic wireless network.  Bose et
al.~\cite{BMSU01} propose GFG-based depth-first face exploration to
implement geocasting. In principle, this solution is stateless,
however Casteigts et al.~\cite{CNS10} point out that it requires
considerable message overhead and the only way to mitigate this
overhead is to pre-process the network topology to give devices
additional information. This preprocessing is stateful and may require
significant communication resources.

Thus, existing geometric geocasting algorithms are either stateful or
so inefficient that their statelessness is ineffectual.

\ \\ \textbf{Our contribution.}  We present new stateless geocasting
routing algorithms.  We describe a stateless controlled flooding
algorithm, SF, which obviates the need for a locally stored
information to prevent multiple retransmissions.  This algorithm is of
independent interest, as it allows to render existing work based on
controlled flooding~\cite{BMSU01,LNLC06,SH06} stateless as well.
Then, we present a stateless concurrent geometric routing algorithm,
SPG, with better scalability and message overhead than SF.  We explore
combinations of these algorithms and greedy routing. We formally prove
the algorithms correct, analyze their message complexity and evaluate
their performance through simulation. From our analysis, it follows
that the presented algorithms are message efficient, provide guaranteed
delivery to the geocast region with low latency, and do not rely on
computation intensive preprocessing.

\section{Notation and Definitions}\label{secNotations}

\noindent
\textbf{Wireless network, message transmission, routing algorithms.} A
\emph{wireless network} is a set of computer communication devices
capable of exchanging messages.  The network is represented as a graph
$G = (V, E)$, where $V$ is a set of devices, and $E$ is a set of edges
that connect them. An edge exists between two devices if they can send
messages directly. Two such devices are called \emph{neighbors}.  The
graph is \emph{fixed maximum degree} if there is constant $k$,
independent of network parameters such that each device has at most $k$
neighbors.  The communication is two-directional and the graph is
undirected.  A network is \emph{connected} if there exists a path
between any two devices.

Every device has unique planar coordinates which \emph{embeds} the
graph into the geometric plane.  A \emph{dominating set} is a subset
of $V$ where every device in $V$ is a neighbor of at least one device
in this subset. A \emph{connected dominating set} induces a subgraph
that is connected.

A \emph{routing algorithm} ensures that a message is delivered from
the \emph{source} device to a \emph{target} device. If the source and
the target are not neighbors, the routing algorithm is executed on
intermediate devices to decide as how to route the messages to
targets.

To help with routing, a message carries routing information. We
consider routing algorithms where the amount of information the message
carries is independent of the network size. That is, we are interested
in \emph{constant message size} routing algorithms. This, for example,
precludes a routing algorithm from requesting the message to carry a
complete traveled route. Each message carries two addresses: the
(immediate) \emph{sender}, i.e. the device transmitting the message,
and the (immediate) \emph{receiver}, i.e. the device the message is
being sent to.

\ \\ \textbf{Computations and fairness.} A \emph{step} in a routing
algorithm is the receipt of a message by the receiver device and local
processing of the message according to the algorithm, which may result
in further messages added to the send queue of the device. A step is
\emph{atomic} if it does not overlap with steps at this or other
devices.

Every device has a \emph{send queue} $SQ$ that collects messages to be
sent. A message is transmitted by taking it from the sender's send
queue and transferring it to the receiver where it is processed
according to the routing algorithm in a single atomic step.

\emph{Computation} is a sequence of atomic steps that starts in an
initial state of the algorithm. A computation is \emph{fair} if every
message that is in a send queue $SQ$ of some device is eventually
either transmitted or removed from this queue during this
computation. That is, a message may not ``get stuck'' in a send queue
forever. To reason about a routing algorithm, we consider its fair
computations. A computation is \emph{finite} if it has a finite number
of steps. A routing algorithm is \emph{terminating} if all its
computations are finite. A terminating routing algorithm never leaves
messages indefinitely circulating in the network.

\ \\ \textbf{Statelessness.}  A routing algorithm is
\emph{stateless} if it is designed such that devices store no
information about messages between transmissions. It is
\emph{stateful} otherwise.

\ \\ \textbf{Flooding.} One of the simplest routing algorithms is
flooding. In \emph{flooding}, the source device sends a message to all
its neighbors. When a device receives this message, it subsequently
sends the message to all its own neighbors. This simple algorithm
guarantees delivery to all devices connected to the source.

If a message is flooded, it may travel over multiple paths. Thus, a
single device may receive the same messages multiple times. To avoid
endless retransmission of messages, flooding must have a mechanism of
eliminating duplicates.  In classic flooding, each device maintains a
flag for each transmitted message. If the message is already
transmitted, and it is received again, the duplicate is discarded.
That is, classic flooding is stateful. In this paper, we present a new
stateless flooding algorithm.

\ \\ \textbf{Planarity, face traversal, mates.}  Simple flooding
requires all devices in the network to transmit the message. This may
not be efficient. Graph planarization offers a way to design more
efficient algorithms.  A graph embedding is \emph{planar} if graph
edges intersect only at vertices. A \emph{connected planar subgraph}
is a subset of vertices and their induced edges such that the
resultant graph is planar and connected. In general, finding a planar
subgraph is a complex task. However, for certain graph classes it is
relatively simple.

A graph is \emph{unit-disk} if a pair of vertices $a$ and $b$ are
neighbors if and only if the distance between them is no more than
$1$. Such graph approximates a wireless network. In such a graph, a
connected planar subgraph may be found by local computation at every
device using Relative Neighborhood or Gabriel Graph
~\cite{BMSU01,GS69,KK00,T80}. Moreover, a local computation on a
unit-disk graph may yield a fixed maximum degree connected dominating
set subgraph~\cite{WAF02}. In our message complexity estimations and
in our simulation, we consider the original graph to be unit-disk.

\emph{Face} is a region on the plane such that, for any two points in
the region, there is a continuous line that connects them without
intersecting graph edges. Note, for example, faces $F$ and $G$ in
Figure~\ref{figSPGExample}.  A planar embedding of a finite graph
divides the plane into a finite set of faces. There areas of each face
but one are finite. They are \emph{internal} faces. One face is an
infinite \emph{external} face.

In a planar graph, messages are routed by traversing such faces using
right- or left-hand-rule. In the \emph{right-hand-rule}, if device $a$
receives a message from device $b$, device $a$ forwards the message to
device $c$ that is nearest to $b$ clockwise. In the
\emph{left-hand-rule}, the message is forwarded to the nearest
counter-clockwise neighbor.  Two messages are \emph{mates} if the
sender of each message is the receiver of the other.  For planar
traversal algorithms, mates also must have the opposite traversal
direction: right- or left-hand-rule.

\ \\ \textbf{Geocasting.}  The problem of \emph{geocasting} is
communicating a message from a source device to all devices located in
a designated \emph{geocast region}. In other words, every device in
the geocast region is a target.  The geocast region is often a circle
or rectangle.  Note that the source itself may be in the geocast
region.  The problem is complicated by the fact that devices in the
geocast region may only be connected through the outside device. See,
for example, devices $f$ and $i$ in Figure~\ref{figSPGExample}. Thus,
message delivery to all devices in the geocast region requires
exploring these outside connecting paths.

In this paper, we present a stateless geocasting algorithm and
its combination with stateless flooding.

\section{Algorithm Descriptions}\label{secDescriptions}

\begin{figure}[htb]
\begin{minipage}[t]{0.5\linewidth}
\begin{tabbing}
1234\=1234\=1234\=1234\=12345\=12345\=12345\=12345\=12345\=12345\=\kill
\textbf{device} $s$\\
\textbf{foreach} $n\in N$ \textbf{do} \\
\>\textbf{add} $M(s,n)$ to $SQ$\\
\\
\textbf{device} $n$\\
\textbf{if} \textbf{receive} $M(a,n)$ \textbf{then}\\
\>\textbf{if} $M(n,a) \in SQ$ \textbf{then} \\
\>\>/* found mate */\\
\>\>\textbf{discard} $M(n,a)$ \textbf{from} $SQ$\\
\>\textbf{else} \\
\>\>\textbf{foreach} $m \in N : m \neq a$ \textbf{do}\\
\>\>\>\textbf{add} $M(n,m)$ to $SQ$\\
\end{tabbing}
\caption{SF pseudocode.} \label{figSFCode}
\end{minipage}
\begin{minipage}[t]{0.5\linewidth}
\begin{tabbing}
1234\=1234\=1234\=1234\=12345\=12345\=12345\=12345\=12345\=12345\=\kill
\textbf{device} $s$\\
/* let $F$ be a face bordering $s$ \\
\>and intersecting the $sr$-line */\\ 
\textbf{add} $L(s,d,F)$ to $SQ$\\
\textbf{add} $R(s,d,F)$ to $SQ$\\
\\
\textbf{device} $n$\\
\textbf{if} \textbf{receive} $L(s,d,F)$ \textbf{then}\\
\>\textbf{if} $R(s,d,F) \in SQ$ \textbf{then} \\
\>\>/* found mate */\\
\>\>\textbf{discard} $R(s,d,F)$ \textbf{from} $SQ$\\
\>\textbf{else} \\
\>\>\textbf{if} $F$ is a juncture \textbf{then}\\
\>\>\>\textbf{foreach} $F'\neq F$ that is\\
\>\>\>\>\>a juncture \textbf{do}\\
\>\>\>\>\textbf{add} $L(s,d,F')$ to $SQ$\\
\>\>\>\>\textbf{add} $R(s,d,F')$ to $SQ$ \\
\>\>\textbf{add} $L(s,d,F)$ to $SQ$ \\
\textbf{if} \textbf{receive} $R(s,d,F)$ \textbf{then} \\
\>/* handle similar to $L(s,d,F)$ */
\end{tabbing}
\caption{SPG pseudocode.} \label{figSPGCode}
\end{minipage}
\end{figure}

\begin{figure}[htb]
\centering
\subfigure[initial step]{\label{figSPG1}
\epsfig{figure=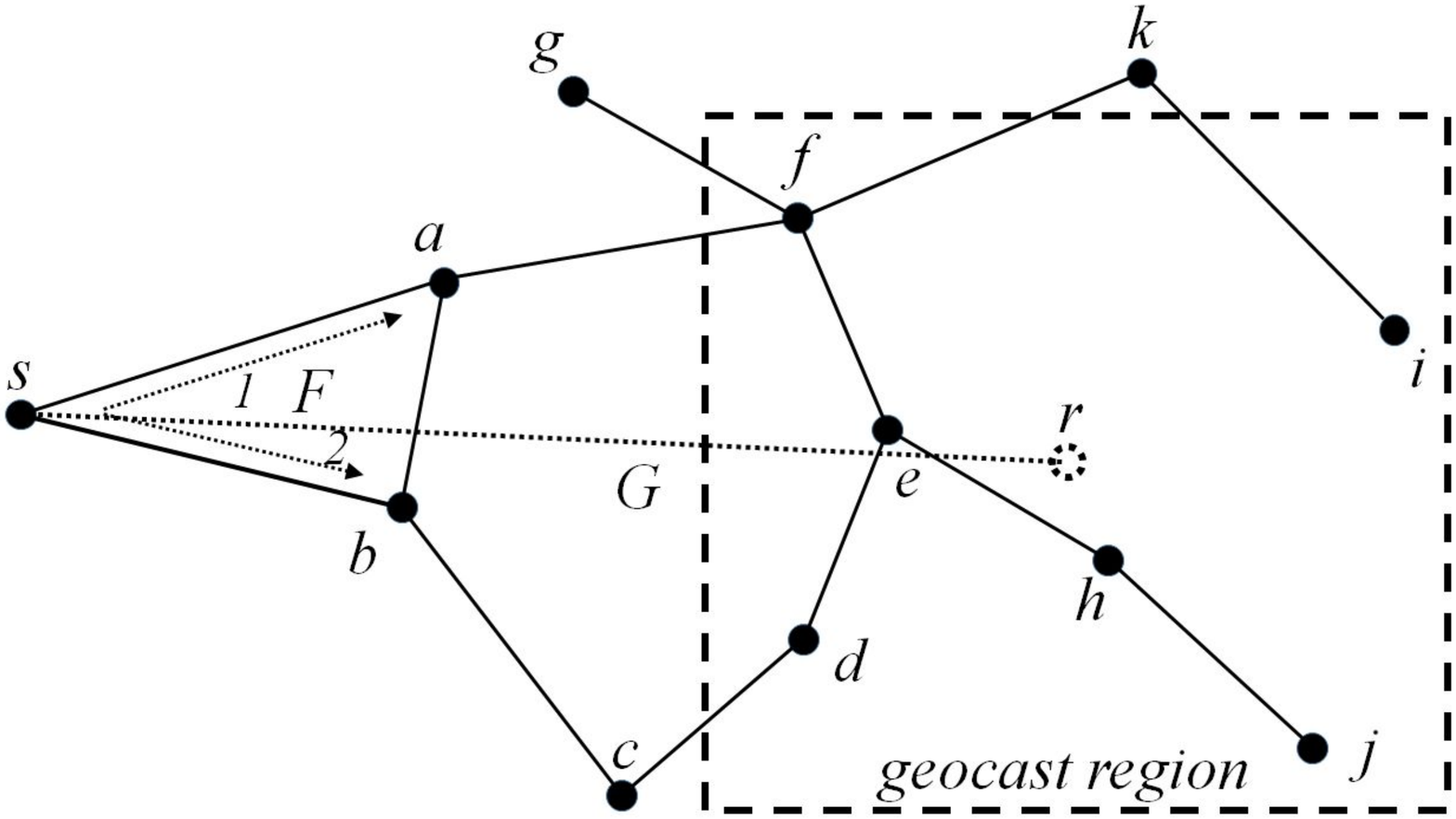,width=6.5cm,clip=}}
\subfigure[mid-computation]{\label{figSPG2}
\epsfig{figure=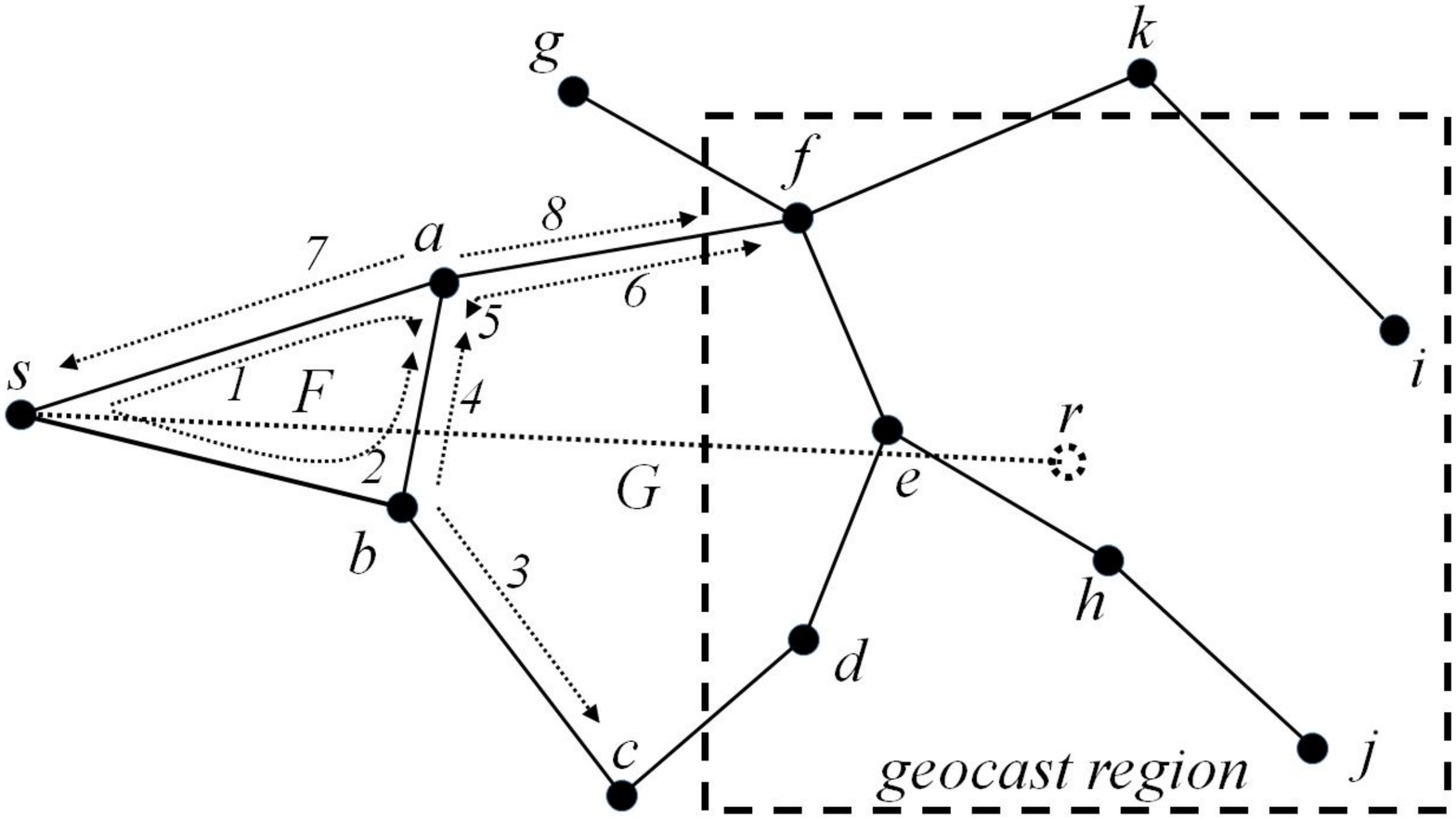,width=6.5cm,clip=}}
\subfigure[completed computation.]{\label{figSPG3}
\epsfig{figure=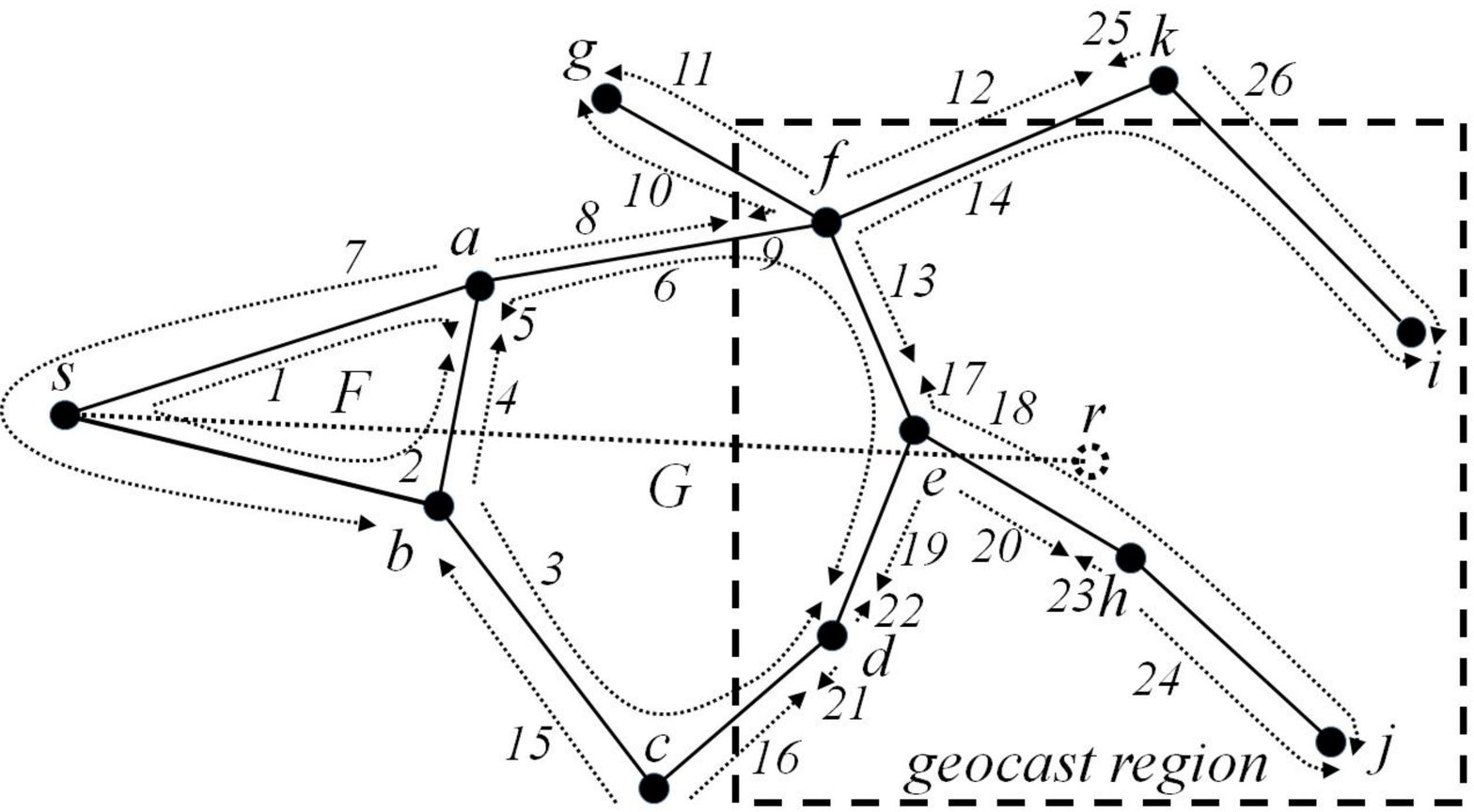,width=6.5cm,clip=}}
\caption{Example computation of SPG.}
\label{figSPGExample}
\end{figure}

\noindent
\textbf{SF.} The pseudcode for \emph{stateless flooding} (SF) routing
algorithm is shown in Figure~\ref{figSFCode}. The algorithm is as
follows.  The source device adds a message $M(sender, receiver)$ to
its send queue $SQ$ to be sent to all devices in its neighbor set $N$.
When a device receives a message from neighbor $a$, it first checks its send
queue for a mate. If a mate exists, both messages are
discarded. Otherwise, the device sends the message to all neighbors
except $a$.

\ \\ \textbf{SPG.} \emph{Stateless planar geocasting} (SPG) algorithm
uses face traversal to limit the number of messages sent during mobile
geocasting.  Let us start the algorithm description with a couple of
definitions. Every message carries the coordinates of the source and
the parameters of the geocast region so it can compute the
\emph{sr-line} --- the source-region line that connects the source
device with the center of the geocast region. A device is a
\emph{juncture} if it is incident to an edge which intersects the
sr-line or the geocast region.  Note that all devices inside the
geocast region are junctures.  A juncture, as any device, is adjacent
to a number of faces. The number of these faces is equal to the number
of the neighbors. For example, in Figure~\ref{figSPGExample}, device
$f$ is adjacent to faces $afe$, $efk$, $kfg$ and $gfa$. Note that some
of these local faces may globally be the same face. For example,
$afg$, $gfk$ and $kfe$ are, in fact, the external face. When receiving a
message traversing a face, a juncture device may determine which face
the message is traversing if the message carries its sender and its
traversal direction: right- or left-hand-rule. For example, if $f$
receives a right-hand-rule message from $a$, then $f$ is able to
determine that the message traverses face $afe$. To simplify the
presentation of SPG, we just refer to a particular face that the
message is traversing.  In SPG, messages carry the source coordinates
$s$, region coordinates $r$ and the message traversal direction $R$ or
$L$.

The pseudocode for SPG is shown in Figure~\ref{figSPGCode}.  The SPG
algorithm operates as follows. The source device sends two messages in
the opposite traversal directions along the face that intersects the
sr-line.  When a device receives a message, it checks its send queue
for a mate. If a mate is present, both messages are discarded.
Otherwise, the device forwards the message along its current face. If
the device is a juncture and the face intersects the sr-line or
geocast region, then the device \emph{splits} the message by sending a
pair of messages in every other face that intersects the sr-line/geocast
region.

We illustrate the operation of SPG with an example shown in
Figure~\ref{figSPGExample}. Device $s$ geocasts a message by sending a
left-hand-rule message $1$ to device $a$ and right-hand-rule message
$2$ to device $b$. See Figure~\ref{figSPG1}. This initiates the
traversal of face $F$. Both $a$ and $b$ are junctures. Device $a$ has
adjacent edges that intersect $sr$-line and the geocast region. Device
$b$ has an adjacent edge intersecting $sr$-line. Once $2$ reaches $b$,
it forwards it to $a$ and splits it by sending messages $3$ and $4$ in
face $G$. See Figure~\ref{figSPG2}. Note that face $sbc$ does not
intersect $sr$-line or the geocast region so no messages are sent
there. When $1$ reaches $a$, it forwards it to $b$ by adding it to its
send queue. Device $a$ also splits $1$ by sending $5$, $6$, $7$ and
$8$.  Once $2$ is received by $a$, it meets its mate in $SQ$ and both
messages are discarded completing the traversal of face $F$. This
computation continues until all messages are delivered to targets. The
result of the complete computation is shown in Figure~\ref{figSPG3}.

\ \\ \textbf{SF+SPG.} For routing, pure SPG uses the planar
subgraph. However, this eliminates the non-planar edges that might be
effective in message transmission. This elimination is unavoidable
outside the geocast region to guarantee delivery to all the
targets. However, inside the geocast region, SPG may be supplemented
by stateless flooding.

Combined algorithm, SF+SPG, uses SPG to route toward and around the
geocast region, and SF to flood inside the region. Each message
carries a mode: flood or planar, and is routed using SF or SPG,
respectively.

Devices outside the geocast region receive and send messages only in
planar mode.  When a device inside the geocast region receives a
message from neighbor $b$, it sends a single flood message to all
neighbors inside the region, and a pair of planar messages with
opposite traversal directions to all neighbors outside the region,
except $b$. If the received message was in planar mode, it is sent
back to $b$, and discarded otherwise.

\ \\ \textbf{SF+SPG+G}. Algorithm SF+SPG may be further combined with
greedy routing to decrease the number of required message
transmissions.  Rather than start SPG at the source device, the
message may be initially transmitted using greedy routing by
transmitting a single message to the neighbor that is closest to the
center of the geocast region. The algorithm switches to SPG only when
the greedy routing encounters local minimum: a device with no neighbors
closer to the geocast region; or when the greedy message actually
reaches the geocast region.

\section{Correctness Proofs and Efficiency Bounds}\label{secProof}

\noindent
\textbf{Correctness proofs.} We focus on SF first. Let us introduce
notation we use for the proofs. A device is \emph{visited} if it
receives the message at least once. An edge is \emph{used} if the
message was sent over it at least once. It is \emph{unused} otherwise.
A visited device is a \emph{border} if it has an adjacent unused link.
A visited device that is not a border is \emph{internal}.

\begin{lemma}\label{lemSFBorder} In SF, every border device holds
a message in $SQ$ to be sent over every unused link and it never holds
a message to be send over a used link.
\end{lemma}

\Proof We prove the lemma by induction. The source sends messages over
the links to its neighbors. Therefore, right before the transmission,
the source is a border device with every link unused and a message to
transmit over this link. Therefore, the conditions of the lemma
hold. Assume the conditions hold at some step of a computation. Let us
consider the next step: a transmission of the message from device $a$
to device $b$.  Device $b$ may be visited or not visited. If $b$ is
not visited, then all its links, except for link to $a$, are
unused. When $b$ receives a message from $a$, it becomes a border
device and it sends messages to all neighbors except $a$. This
satisfies the conditions of the lemma. If $b$ is already visited, then
it has a message to be sent to $a$. This massage is a mate of the
message received by $b$ from $a$. By the algorithm, these two messages
are discarded. That is, once the message is transmitted to a visited
device and uses the channel, there are no messages to be sent over
this used channel. Again, the conditions of the lemma hold.  \QED

\begin{theorem}
SF guarantees termination and delivery from the source to all target
devices connected to the source.
\end{theorem}

\Proof Once the source device has a message to send, it sends to all its
neighbors. That is, it becomes a border device. According to
Lemma~\ref{lemSFBorder}, every border device has a message to transmit
over unused channels.  Since we consider fair computations of routing
algorithms, this message is eventually going to be transmitted. If the
receiver device is not visited, it becomes visited and sends messages
to all its neighbors. Eventually, all devices connected to
the source will be visited, and all channels used. That is, SF
delivers the message to all devices connected to the source.  Note
that according to Lemma~\ref{lemSFBorder}, once the channel is used,
there are no messages to be sent across it. That is, SF terminates.
\QED

We now prove correctness of SPG. Let us introduce additional
terminology. A device is \emph{segment-visited} if it was visited
during the traversal of this face.  A \emph{visited segment} of a face
is a sequence of neighbor devices that have been segment-visited. A
\emph{segment-border} of a visited segment is a segment-visited device
that has an edge adjacent to this face that has not been used. Note
that an edge is adjacent to two faces. Thus for SPG, an edge may be
used in one face and not used in the other. A visited device that is
not a border is \emph{segment-internal}.  Two faces are
\emph{adjacent} if they share a common juncture device, and are
\emph{juncture connected} if there exists a sequence of adjacent faces
from one to the other.

\begin{lemma}\label{lemBorderMessages}
In SPG, every segment-border device always holds a message to be sent
over unused adjacent edge. An internal device never holds such
message.
\end{lemma}

\Proof The proof is by induction on the devices of the face having the
visited segment. A visited segment is created when a juncture device
is visited. This juncture may be the source device $s$ or another
juncture splitting the message when it is visited in an adjacent
face. Once the visited segment is created, it contains a single border
device with two messages sent in the opposite traversal
directions. This is our base case. Let us consider a computation of
SPG where every segment of every face is as stated in the conditions of the
lemma. Let us focus on a particular face $F$ and a message transmission
affecting its segments.

First, let us consider a message transmission by a device adjacent to
$F$. It may only be a border device. The message recipient may be a
non-visited device or a border of another visited segment. If the
recipient is a non-visited device, once the message is received, the
recipient forwards the message to its neighbor. That is, the recipient
becomes a new border device with the sent message while the sender becomes an
internal device without a message. Thus, the conditions of the lemma are
satisfied. If the recipient is a border device of an adjacent segment,
by the induction hypothesis, the recipient holds a mate to be
transmitted to the original sender. The two messages are discarded and
the two adjacent segments merge preserving the conditions of the
lemma.

Let us now contemplate a message transmission by the device that is not
adjacent to $F$. The only way it may affect $F$ is if the transmission
is to a juncture of $F$ in an adjacent face. However, by the design of
the algorithm, the juncture is instantly visited in every adjacent
face. Hence, the message transmission should encounter a border device
with a mate and be eliminated.

That is, regardless of the kind of message transmissions we consider,
the conditions of the lemma are preserved.  \QED

\begin{lemma}\label{lemAllFaceVisited}
In SPG, if a face has a visited segment, every device adjacent to this
face is eventually visited.
\end{lemma}
\Proof If a face contains a non-visited device, then at least one
non-visited device is adjacent to a border device of a visited
segment. According to Lemma~\ref{lemBorderMessages}, this border device
has a message to be sent to the non-visited adjacent device.  Since we
only consider fair computations of  routing algorithms, this message is
eventually transmitted. Once the message is transmitted, the adjacent
device becomes visited. This process continues until all devices are
visited.  \QED

\begin{lemma}\label{lemAllVisited}
In SPG, every device in a face connected to the source device face is
eventually visited.
\end{lemma}

\Proof We start with the face that contains the source device $s$. The
source device is in a visited segment.  According to
Lemma~\ref{lemAllFaceVisited}, every device in this face is eventually
visited. By the design of the algorithm, a juncture device is
instantaneously visited in all its adjacent faces. This means that
visiting every device in the face that contains $s$ creates visited
segments in every face that is adjacent to it. Repeated application of
Lemma~\ref{lemAllFaceVisited} proves this lemma.  \QED

\begin{proposition}\label{propAllConnected}
In a planar graph, if a target device is connected to the source
device, then this target device lies on a face connected to the source
device face.
\end{proposition}

The below theorem follows from Proposition~\ref{propAllConnected} and
Lemma~\ref{lemAllVisited}.

\begin{theorem}
SPG guarantees termination and delivery of a message from the source
to all target devices connected to the source.
\end{theorem}

\begin{figure}\label{fig2e}\centering
\epsfig{figure=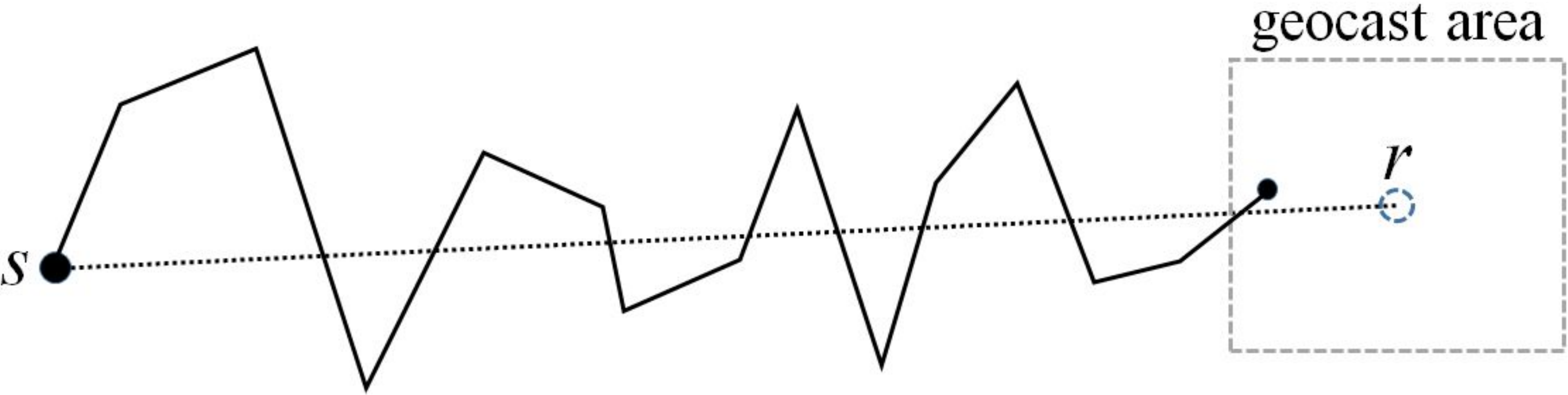,width=7cm,clip=}
\caption{Example graph for $2E$ message cost in SPG.}
\end{figure}

\noindent\textbf{Message costs.} In the case of stateless flooding
(SF), each device sends exactly one message.  That is, the total
number of messages is $E$. In case of stateless planar geocasting
(SPG), a message may be sent along each face. An edge may be adjacent
to two faces. Hence, SPG may send $2E$ messages.  This estimate is
tight. See, for example, Figure~\ref{fig2e} where SF sends $E$
messages while SPG sends $2E$. That is, in the worst case, SPG may be
twice as costly as SF.

However, for most graphs, SPG is significantly more efficient. To give
a more realistic message cost estimate for SPG, we make several
assumptions about the network graphs. First, we assume that the
geocast region is square.

The graph is \emph{face smooth} if there are two constants $c_1$ and
$c_2$ that are independent of network parameters such that (i) for
each face $\rho ^2 < c_1 a$ where $\rho$ is the perimeter of the face,
and $a$ its area, and (ii) for any two points in the graph, $a_s < c_2
\frac{\pi d^2}{4}$ where $a_s$ is the area of all internal faces that
intersect the line between these two points and $d$ is the Euclidean
distance between them. For an internal face, the area computation is
straightforward; for the external face, an area of an arbitrary figure
enclosing the graph, for example convex hull, is considered.  The
first assumption places limits on how "ragged'' the perimeter of the
face may be, while the second limits how ``uneven'' the faces may be
in size by assuming that the area of all intersecting faces is
included in a certain disk whose diameter is related to the distance
between two devices.

\begin{lemma} \label{lemFaceSmoothBound}
For face smooth graphs, the message cost of SPG+SF is less than
\[d \frac{k}{2}\sqrt{\pi c_1 c_2} + k\sqrt{c_1 A} + 2kG +
    2k \sqrt{\pi c_1 c_2 G}, \]
where $d$ the length of $sr$-line, $A$ and $G$ are the respective
areas of the whole graph and the geocast region, $k$ is the maximum
degree and $c_1$ and $c_2$ are constants independent of the graph
parameters.
\end{lemma}

\Proof We arrive at the estimate by bounding the number of messages it
takes to carry out individual parts the algorithm: to traverse the
faces (both internal and external) intersecting $sr$-line, to flood the
geocast region, and to traverse the faces that intersect it. Note that
due to the design of the algorithm, the external face is traversed only
once. Hence, we account for the messages it takes to traverse it only
once as well.

For the $sr$-line, combining the two assumptions for the face smooth
graphs, we get:
\[\rho_s^2 < c_1 c_2 \frac{\pi d^2}{4}, \] where $\rho_s$
is the sum of all perimeters of internal faces intersecting the
$sr$-line.  If $k$ is the maximum degree of the graph, the maximum
number of devices lying along the perimeter $\rho_s$ is
$k\rho_s$. Hence, the number $m_{sr}$ of messages sent traversing
these faces is:
\[ \left(\frac{m_{sr}}{k}\right)^2 < c_1 c_2 \frac{\pi d^2}{4}  \]

That is,
\[ m_{sr} < d \frac{k}{2}\sqrt{\pi c_1 c_2}\]

Similarly, if SPG+SF intersects the external face, the number of
messages $m_e$ it takes to traverse it can be bounded as follows:
\[ m_e < k\sqrt{c_1 A}\]

\begin{figure}\centering
\epsfig{figure=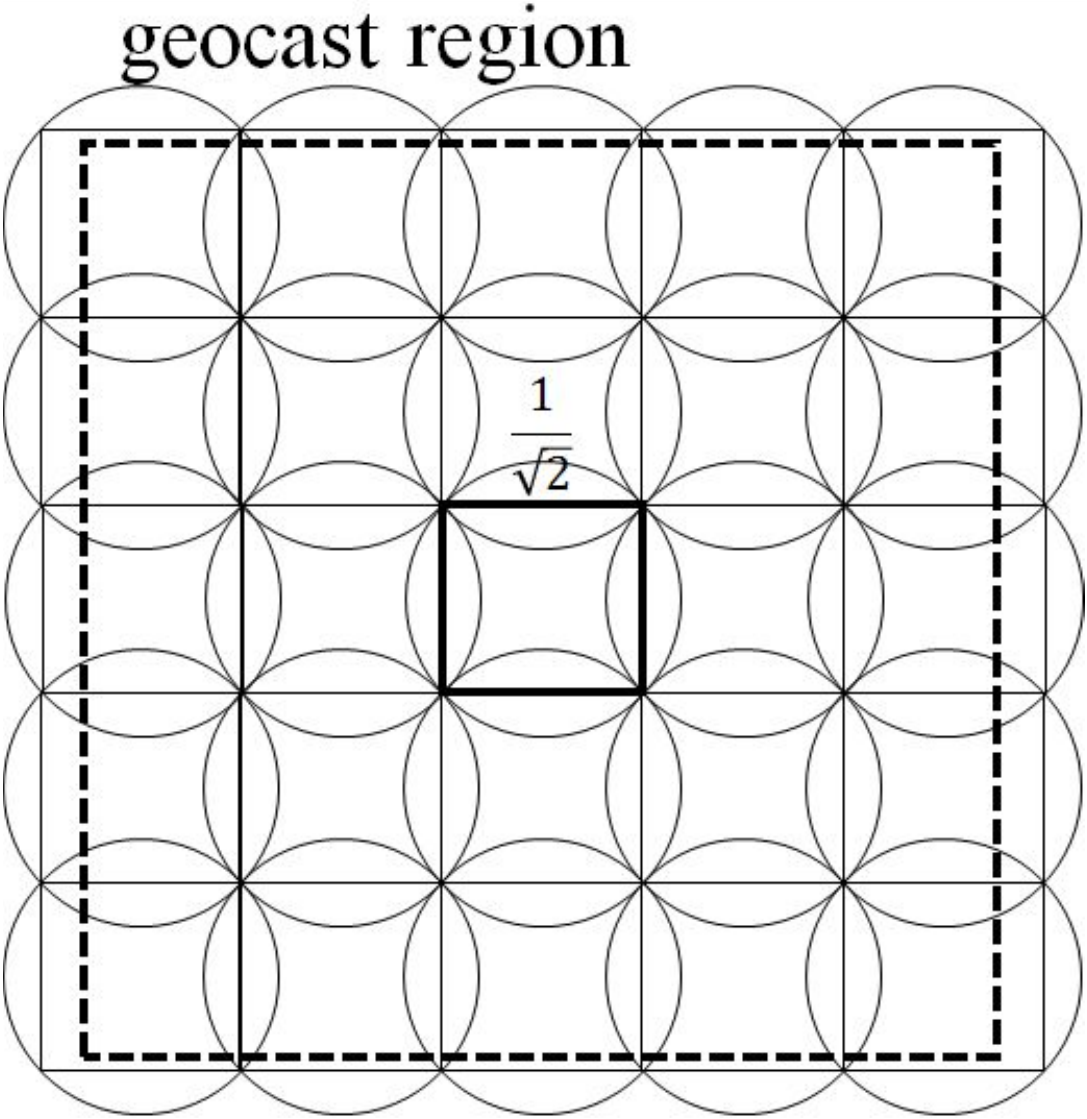,width=7cm,clip=}
\caption{Estimating the number of devices in the geocast region.}
\label{figG}
\end{figure}

Let us estimate the number of messages expended to deliver to all the
devices in the geocast region $G$.  We first bound the number of messages
sent by the flooding part of the algorithm. We completely cover the
whole geocast region with unit-disks as shown in Figure~\ref{figG}. In
this arrangement, each unit disk covers a square with side length of
$\frac{1}{\sqrt{2}}$. With $k$ being the maximum device degree, the
number of devices in the geocast region, and therefore, the number of
messages $m_f$ it takes to flood it, is:
\[ m_f < k \frac{G}{\left(\frac{1}{\sqrt{2}}\right)^2} = 2kG \]

Let us now estimate the number of messages sent across faces
intersecting the geocast region. Since the area is square, the length of
its side is $\sqrt{G}$. By the second assumption of the face smooth
graph, the area of all internal faces that intersect this side is less
than $c_2 \frac{\pi G}{4}$. By the first assumption, the perimeters of
these faces is less than $\sqrt{c_1 c_2 \frac{\pi G}{4}}$. Taking into
account that the geocast region has four sides and that the maximum device
degree is $k$, for the bound of the number of messages $m_f$ on the
faces that intersect the geocast region is as follows:

\[m_g < 4k\sqrt{c_1 c_2 \frac{\pi G}{4}} =  2k \sqrt{\pi c_1 c_2 G} \]
Adding the bounds for individual parts of the message estimate
$m_{sr}$, $m_e$, $m_f$ and $m_g$, we obtain the bound of the lemma.
\QED

The following theorem follows from Lemma~\ref{lemFaceSmoothBound}.

\begin{theorem}\label{trmBound} For face smooth graphs of bounded degree,
if the geocast region size is constant, the message cost for SPG+SF is
in $O(d + \sqrt{A})$, where $d$ is the length of the $sr$-line and $A$
is the area covered by the graph.
\end{theorem}

Let us compare this bound with the message complexity of ordinary
flooding. If the number of devices in the graph is proportional to this
area, the message cost of flooding is in $\Omega(A)$. In other words,
the message cost of SPG+SF is proportional to the linear dimensions of
the geocast region while the cost of flooding is quadratic.

\section{Simulation}
\noindent
\textbf{Setup.} In their classic study of unicast geometric routing
algorithms, Kuhn et al~\cite{KWZZ03} use a particular simulation setup
to thoroughly evaluate the performance of their algorithms. We extend
the setup similar to theirs to use in our simulation.

Specifically, we populate a $10 \times 10$ unit square field with
devices placed uniformly at random to achieve a specific network
density.  The total number $n$ of devices is equal to the area of the
field divided by area of the unit circle and multiplied by the
required density $d$. That is $n = d\frac{100}{\pi}$. We randomly pick
the source device and randomly place a square geocast region so that
it fits into the field completely.  We then calculate each device's
neighbors as follows. We first construct a unit-disk graph. For the
planar geocasting algorithms, we also compute Gabriel subgraph and
connected dominating set on it.

\emph{Experiment} is a single delivery of a message from a particular
source to a particular geocast region. In other words, it is a single
complete computation of an algorithm. For each experiment, we generate
a new random graph with a randomly selected source and randomly placed
geocast region. For each specific data point, we conduct 1000
experiments.

\begin{figure}
\centering
\subfigure[by network density]{\label{figOverheadByDensity}
\epsfig{figure=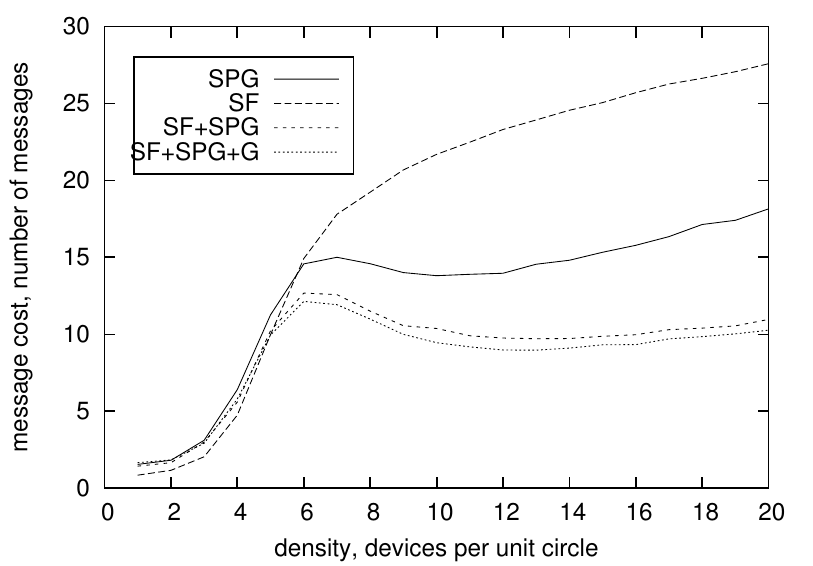,width=0.48\linewidth,clip=}}
\subfigure[by geocast region size, length of geocast square side]{\label{figOverheadByArea}
\epsfig{figure=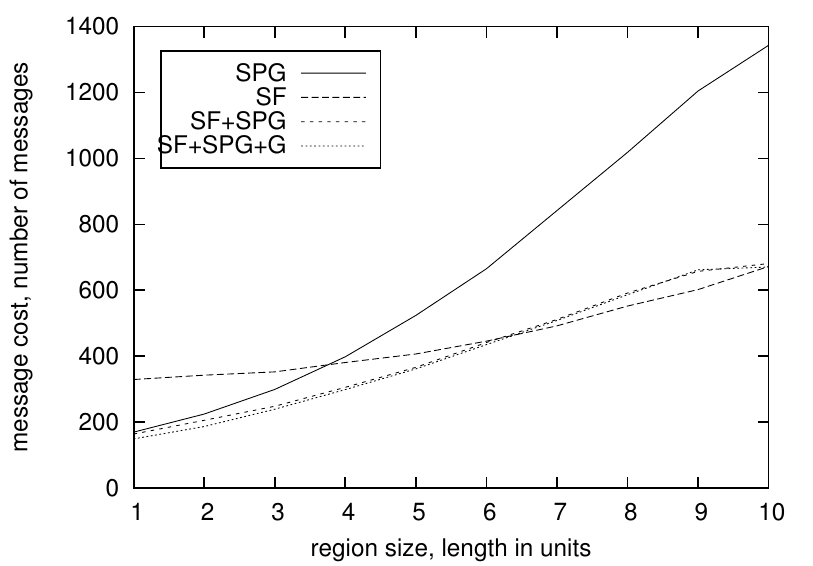,width=0.48\linewidth,clip=}}
\subfigure[by network size, length of field square side]{\label{figOverheadByScale}
\epsfig{figure=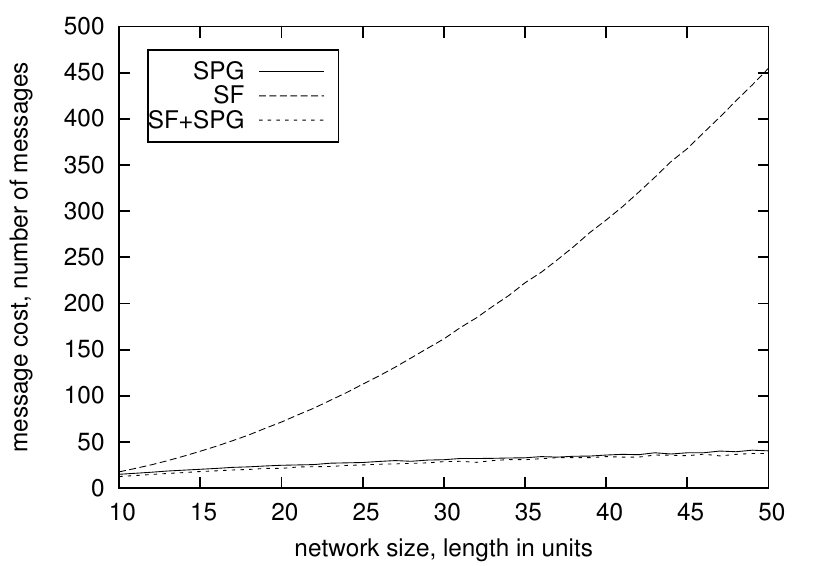,width=0.48\linewidth,clip=}}
\caption{Message cost normalized to geocast region size.}
\label{figOverhead}
\end{figure}

\begin{figure}
\centering
\subfigure[by network density]{\label{figLatencyByDensity}
\epsfig{figure=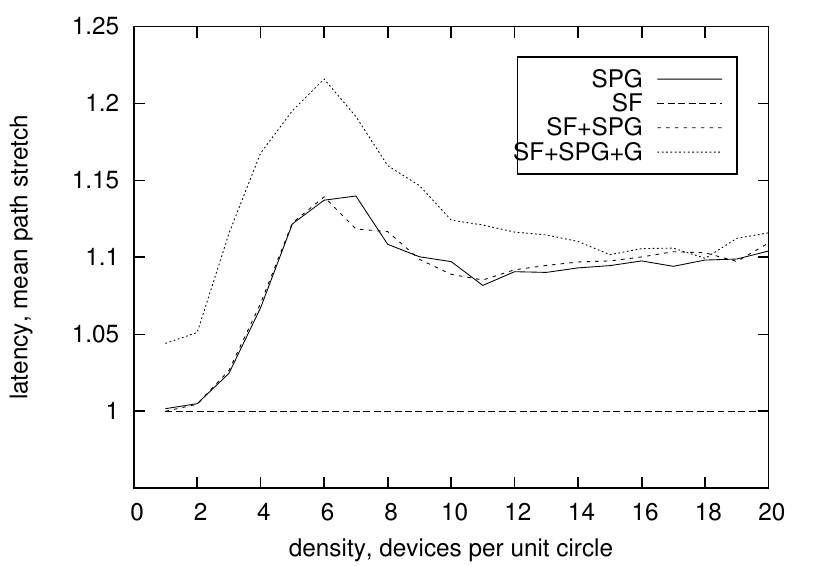,width=0.48\linewidth,clip=}}
\subfigure[by geocast region size]{\label{figLatencyByArea}
\epsfig{figure=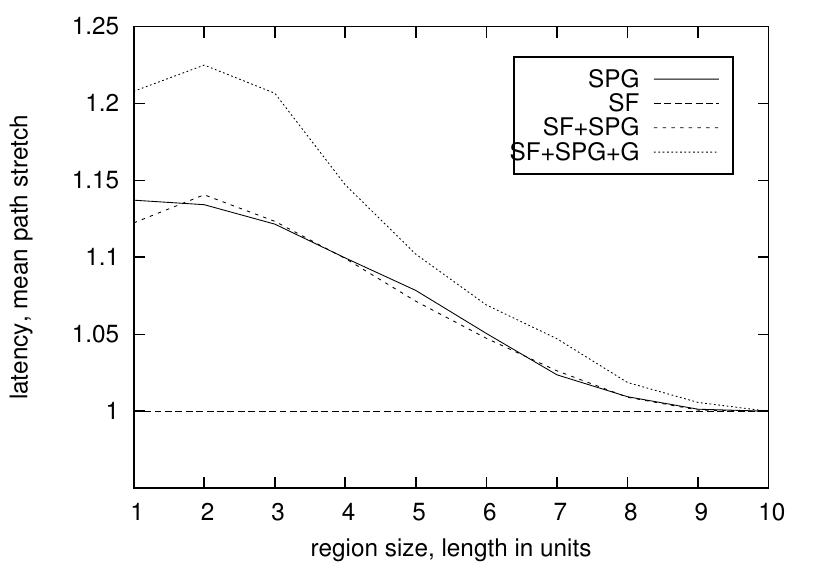,width=0.48\linewidth,clip=}}
\caption{Arrival latency normalized to optimal path.}
\label{figLatency}
\end{figure}

\noindent
\ \\ \textbf{Results.} We evaluate the algorithms on message cost and
latency.  \emph{Message cost} is the number of messages it takes to
deliver to all devices in the geocast region. Message cost
quantifies the amount of network resources necessary to deliver the
message.  \emph{Latency} of message arrival is the shortest path taken
by the algorithm to reach the device in the geocast region that is
furthest away from the source. Devices not connected to the source are
not counted.  \emph{Path stretch} is latency normalized to the optimal
path to this device.  Latency quantifies the time it takes to deliver
the message to every device in the geocast region.

We estimate message cost by varying three parameters: network density,
geocast region size and complete communication field size. When we
vary one of the three parameters, the other two are held constant at:
$7$ devices per unit square for density, $3\times 3$ units for geocast region
size and $10\times 10$ units for field size.  Figure~\ref{figOverhead}
presents the simulation results.

We study latency by varying network density and geocast region size.
The results for latency are shown in Figure \ref{figLatency}.

\ \\ \textbf{Analysis.} The results of the message cost study are
intuitive. SF becomes comparatively costly as the density of the
network increases (see Figure~\ref{figOverheadByDensity}). Indeed, SF
delivers the message to every device in the whole network, regardless
of whether they are inside or outside the geocast region. The delivery
to the outside devices is overhead.  As the density grows, the
ratio of outside devices to inside devices also grows. The overhead
grows with the increase of this ratio.  SF+SPG performs better than
either of the two individual algorithms. The combined algorithm
achieves message savings compared to pure SPG since it floods the
geocast region. When flooding, the algorithm sends only one message
per edge, while SPG may potentially send two messages. Adding greedy
to the combined algorithm helps further reduce message cost.

Let us consider geocast region variation. Again, since SF sends one
message per edge, while SPG may potentially send two messages,
SF outperforms SPG as the geocast region size approaches field size
(see Figure~\ref{figOverheadByArea}).  The growth of the field size
adversely affects SF's performance (see
Figure~\ref{figOverheadByScale}).

The results of the latency study are also intuitive for the most part.
SF is always latency-optimal as all possible paths are traveled.  The
other algorithms achieve similar mean path stretch, whether under
varied density or geocast region size. However, adding a greedy
component dramatically worsens the algorithm's performance: greedy
routing does not have the advantage of concurrently exploring multiple
paths to find the faster one to deliver the message.

\section{Future Research}
In conclusion, we would like to point out several ways the algorithms
presented in this paper may be extended. We assumed fault-free
delivery for our geocasting algorithms. It would be interesting to
incorporate fault-tolerance and message-loss resilience into them.

Other practical considerations enhance the applicability of our
algorithms. For example, both SF and SPG, upon arrival of a message to
a device, require examination of the send queue at this device. This
examination has to be implemented at all levels of the network stack
of each device. Such cross-stack queue examination presents an
interesting implementation challenge.

\bibliographystyle{plain}


\end{document}